\documentclass[twocolumn,trackchanges]{aastex61}
\usepackage{natbib} 
\usepackage{hyperref}
\usepackage{xcolor}
\usepackage{amsmath}
\usepackage{rotating}
\usepackage{array}
\usepackage{bibentry}
\usepackage{amsmath}
\usepackage{enumitem}  
\usepackage{booktabs}
\usepackage{multirow}
\usepackage{gensymb}
\defcitealias{ag2017}{G17}

\newcommand{\hiz}{high-$z$}

\newcommand{\mum}{$\mu$m }
\newcommand{\spit}{\emph{Spitzer}}
\newcommand{\her}{\emph{Herschel} }

\newcommand{\kms}{km s$^{-1}$}

\newcommand{\aco}{$\alpha_{\rm CO}$}
\newcommand{\acou}{$M_{\odot} $(K km s$^{-1}$ pc$^{-2}$)$^{-1}$}
\newcommand{\lu}{K km s$^{-1}$ pc$^{-2}$}

\newcommand{\coff}{CO($5-4$) }
\newcommand{\fgas}{$f_{\rm gas}$}

\shorttitle{High gas fraction in a CO-selected main-sequence galaxy at $z > 3$}
\shortauthors{Gowardhan et al}

\begin{document}

	\title{High gas fraction in a CO-selected main-sequence galaxy at $z > 3$}

	\def\andname{\hspace*{-0.5em}}

\def\andname{\hspace*{-0.5em}}
	
\author[0000-0002-3310-5859]{Avani Gowardhan} 
\affiliation{Department of Astronomy, Cornell University, Ithaca, NY 14853, USA}  
 
\author[0000-0001-9585-1462]{Dominik Riechers}
\affiliation{Department of Astronomy, Cornell University, Ithaca, NY 14853, USA}  

\author[0000-0002-2263-646X]{Riccardo Pavesi}
\affiliation{Department of Astronomy, Cornell University, Ithaca, NY 14853, USA}  

\author[0000-0002-3331-9590]{Emanuele Daddi}
\affiliation{Irfu/Service d'Astrophysique, CEA Saclay, Orme des Merisiers, F- 91191 Gif sur Yvette, France}  

\author[0000-0001-7147-3575]{Helmut Dannerbauer}
\affiliation{Instituto de Astrof\'isica de Canarias (IAC), E-38205 La Laguna, Tenerife, Spain}
\affiliation{Universidad de La Laguna, Dpto. Astrof\'isica, E-38206 La Laguna, Tenerife, Spain}  
 
\author[0000-0002-7176-4046]{Roberto Neri}
\affiliation{Institut de Radioastronomie Millim\'etrique, 300 rue de la Piscine, F-38406, Saint-Martin d’H\'eres, France}

%=============================%
% ABSTRACT
%=============================%

\begin{abstract}

We report NOrthern Extended Millimetre Array (NOEMA) observations of warm molecular gas traced by \coff in a $z \sim 3.2$ gas-rich main-sequence galaxy (MS), initially serendipitously detected in CO($3-2$) emission in `blind' deep NOEMA observations. Our target shows a gas excitation consistent with that seen in $z \sim 1.5$ MS galaxies ($L'_{\rm CO( 5 - 4)}/L'_{\rm CO (3 - 2)} = 0.41 \pm 0.14$), albeit toward the low end, as well as a similar star formation efficiency based on the CO($3-2$) line luminosity and the $L_{\rm IR}$. However, it shows a high molecular gas fraction ($f_{\rm gas} =  0.9\pm 0.2$) as compared to $z\sim 1.5$ MS galaxies ($f_{\rm gas} \sim 0.42$), consistent with a cosmologically increasing gas fraction beyond $z\gtrsim3$ and our current understanding of scaling relations between $z$, $f_{\rm gas}$, the stellar mass $M_*$, and the specific star formation rate sSFR. Our results are consistent with recent findings by the COLDz and ASPECS molecular line scan surveys and suggest that deep searches for CO emission are a powerful means to identify gas-rich, star-forming galaxies at high redshift.

\end{abstract}

%=============================%
% INTRODUCTION
%=============================%

\section{Introduction} Observations of molecular gas - the fuel for star formation - in sizable galaxy samples at high-$z$ are essential to understanding the onset and evolution of the peak epoch of cosmic star formation and stellar mass assembly at $z \sim 1-3$ \citep[see][for a review]{carilli2013}. Star-forming galaxies (SFGs) at all cosmic epochs show a redshift-modulated correlation between the stellar mass and the star-formation rate (SFR) - the galaxy main-sequence - suggesting that the bulk of star formation takes place in quasi-steady state, with galaxies undergoing short-lived starburst activity lying significantly above the galaxy main-sequence at any redshift \citep[e.g.][]{rodighiero2011, speagle2014}. Observations of the molecular gas traced by CO as well as dust-based measurements of the total gas and dust mass suggest that the observed increase in star-formation rates in \hiz$ $ SFGs is driven concurrently by increasing gas fraction (\fgas) and star-formation efficiency (SFE) \citep[e.g.][]{tacconi2013, genzel2015, scoville2016, pavesi2018}. However, while there is a general agreement on the evolution of the molecular gas fraction and specific star formation rate (sSFR) up to $z \sim 2$, there is considerable debate about its evolution beyond that epoch. While some studies find a continuing increase in the molecular gas fraction at $z \gtrsim 3$ \citep{tan2013, dz2015, dz2017}, as expected from theoretical models \citep[e.g][]{obreschkow2009a, lagos2011}, other measurements indicate a plateauing or even a decline of the molecular gas fraction at the highest redshifts \citep[e.g][]{saintonge2013, troncoso2014, bethermin2015,dz2015, schinnerer2016}. CO line stacking of 78 galaxies at a mean redshift of $z\sim 2.4$ also shows a lower molecular gas fraction than expected for massive main-sequence galaxies \citep{pavesi2018}. This disagreement can be attributed to the scarcity of molecular gas detections in MS galaxies at $z \gtrsim 3$. CO detections in SFGs at $z\gtrsim 3$ are currently largely restricted to highly lensed systems (magnified $30-60 \times$ \citealt{coppin2007, riechers2010a, saintonge2013, dz2017}). Searches in unlensed Lyman-Break galaxies (LBGs) at $z\sim 3$ have had limited success, with only two detections to date \citep{magdis2012,tan2013,magdis2017}. 

Observing both low-$J$ and high-$J$ CO lines in high-$z$ SFGs is important as they trace the cold and warm molecular gas phases respectively. While CO Spectral Line Energy Distributions (SLEDs) have been studied in FIR-bright submillimeter galaxies (SMGs) and quasars at \hiz$ $ \citep[e.g.][]{weiss2005, riechers2006, weiss2007, riechers2011b, riechers2011c, danielson2011, riechers2013, bothwell2013, yang2017,strandet2017}, these systems are undergoing intense star-formation, have small gas depletion timescales \citep{yang2017}, and are unlikely to be representative of MS galaxies. The CO SLED has been only sparsely sampled for more `normal' \hiz$ $ star-forming galaxies, with observations limited to four BzK galaxies at $z \sim 1.5$ \citep{daddi2015} and one lensed source at $z\sim 3.6$ \citep{dz2017}. While low-$J$ ($J_{\rm upper} = 1,2,3$) CO line ratios in these systems resemble those of star-forming galaxies in the local universe, CO($5-4$) observations reveal the presence of an additional, warmer molecular gas component, demonstrating the necessity of sampling the CO SLED at multiple $J$s to accurately probe ISM properties \citep{daddi2015}. 

We here present observations of CO($5-4$) emission in EGSIRAC J141912.03+524924.0 (hereafter EGS141912), a gas-rich MS galaxy at $z \sim 3.2$, detected serendipitously in CO($3-2$) emission \citep[][hereafter \citetalias{ag2017}]{ag2017}. Our new observations confirm the target redshift and provide some of the first constraints on the molecular gas excitation and star formation efficiency in $z > 3$ MS galaxies. 

The paper is organized as follows: we present the observations in \autoref{sec:obs} and the spectral energy distribution (SED) fitting in \autoref{sec:results}. In \autoref{sec:discussion} and \autoref{sec:conclusions}, we discuss our results and conclusions. We use a $\Lambda$CDM cosmology, with $H_{0} = 71$ km s$^{-1}$ Mpc$^{-1}$, $\Omega_{\rm M} = 0.27$, and $\Omega_{\Lambda} = 0.73$ \citep{spergel2007}.

%=============================%
% OBSERVATIONS
%=============================%

\section{Observations}\label{sec:obs}
\subsection{CO observations}

NOEMA observations of the CO($5-4$) line ($\nu_{\rm rest} = 576.26793$ GHz) in EGS141912 were conducted in April 2017 (Program ID W16DR), with 8 antennas in the compact D configuration, for a total on-source time of 9.2 hours split across two tracks. Weather conditions were good for both tracks, with a precipitable water vapor (pwv) of $2-15$ mm, with most of the observations taken in good weather. 3C273 was used as the absolute flux calibrator, and the source J1418+546 was used for phase and bandpass calibration. The WideX correlator (bandwidth $\sim 3.6$ GHz) was tuned to a frequency of $136.605$ GHz. Observations were carried out in a dual polarization mode, with a binned spectral resolution of $\sim 2.5$ MHz ($\sim 5.5$ km s$^{-1}$ at 136 GHz). All observations were calibrated using the IRAM PdBI data reduction pipeline in CLIC (Continuum and Line Interferometer Calibration), with subsequent additional flagging by hand. The reduced visibility data were imaged in the software {\tt {MAPPING}}, using the tasks {\tt {UV\_MAP}} and {\tt {CLEAN}}, using natural baseline weighting and the Hogbom cleaning algorithm. The final synthesized beam size is $3.''0 \times 2.''5$. The rms noise in the cube is $1.0$ mJy beam$^{-1}$ per $\sim 15.5$ \kms$ $ channel. Upon binning the line-free channels, we obtain an rms noise of $0.03$ mJy beam$^{-1}$ in the continuum map. 

\subsection{VLA observations}

Radio continuum observations covering EGS141912 have been conducted using the NSF's Karl G. Jansky Very Large Array (VLA) over 3 epochs in July-September 2013 (Program IDs 13B-289 and 13A-449). Observations were made in dual polarization using the X-band receivers in the C and CnB array configurations, with a 2 GHz bandwidth ($7.988 - 9.884$ GHz) sampled at a spectral resolution of 1 MHz. The total on-source time was 2.5 hours. 3C295 and J1419+5423 were used for absolute flux and phase calibration, respectively. 

The VLA reduction pipeline in CASA$v 5.0.0$ was used to flag and calibrate the observations. The weights for the visibilities were calculated using {\tt STATWT} for the reduced measurement sets from each observational epoch, and they were combined into a single measurement set using the task {\tt CONCAT}. The final measurement set was imaged and cleaned using the CASA task {\tt TCLEAN}, using natural weighting to maximize point source sensitivity, and a pixel size of $0.''5 \times 0.''5$. Primary beam correction was applied during the cleaning process. All channels were binned together during cleaning. The resulting cleaned image has an rms noise of 1.3 $\mu$Jy beam$^{-1}$ over the entire 2 GHz bandwidth, and a synthesized beam size of $3''.1 \times 2''.3$ (PA: $-76\degree$). 

%=============================%
% FIGURE 1
%=============================%

\begin{figure*}
\includegraphics[width=0.49\textwidth]{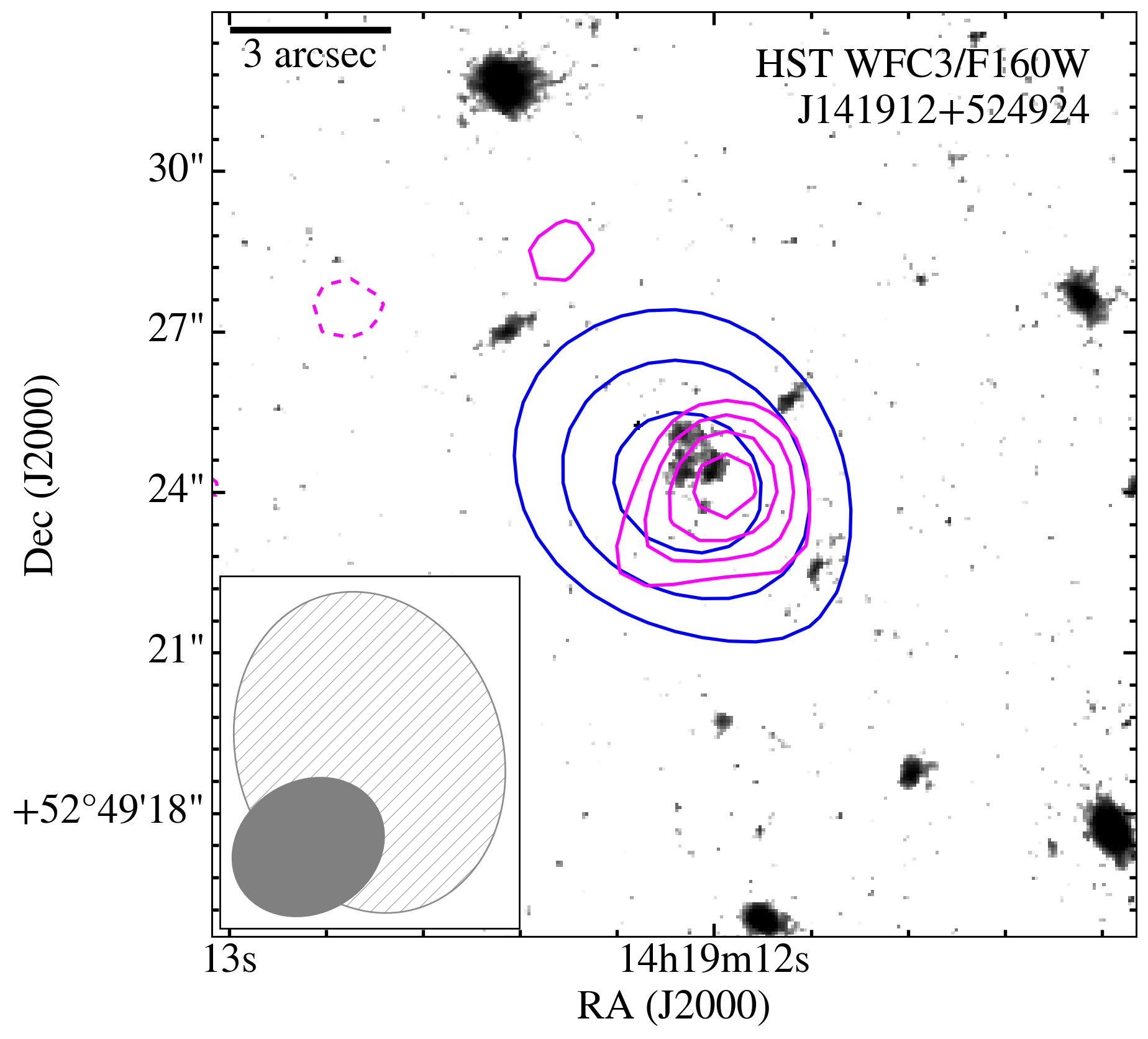}
\includegraphics[width=0.48\textwidth]{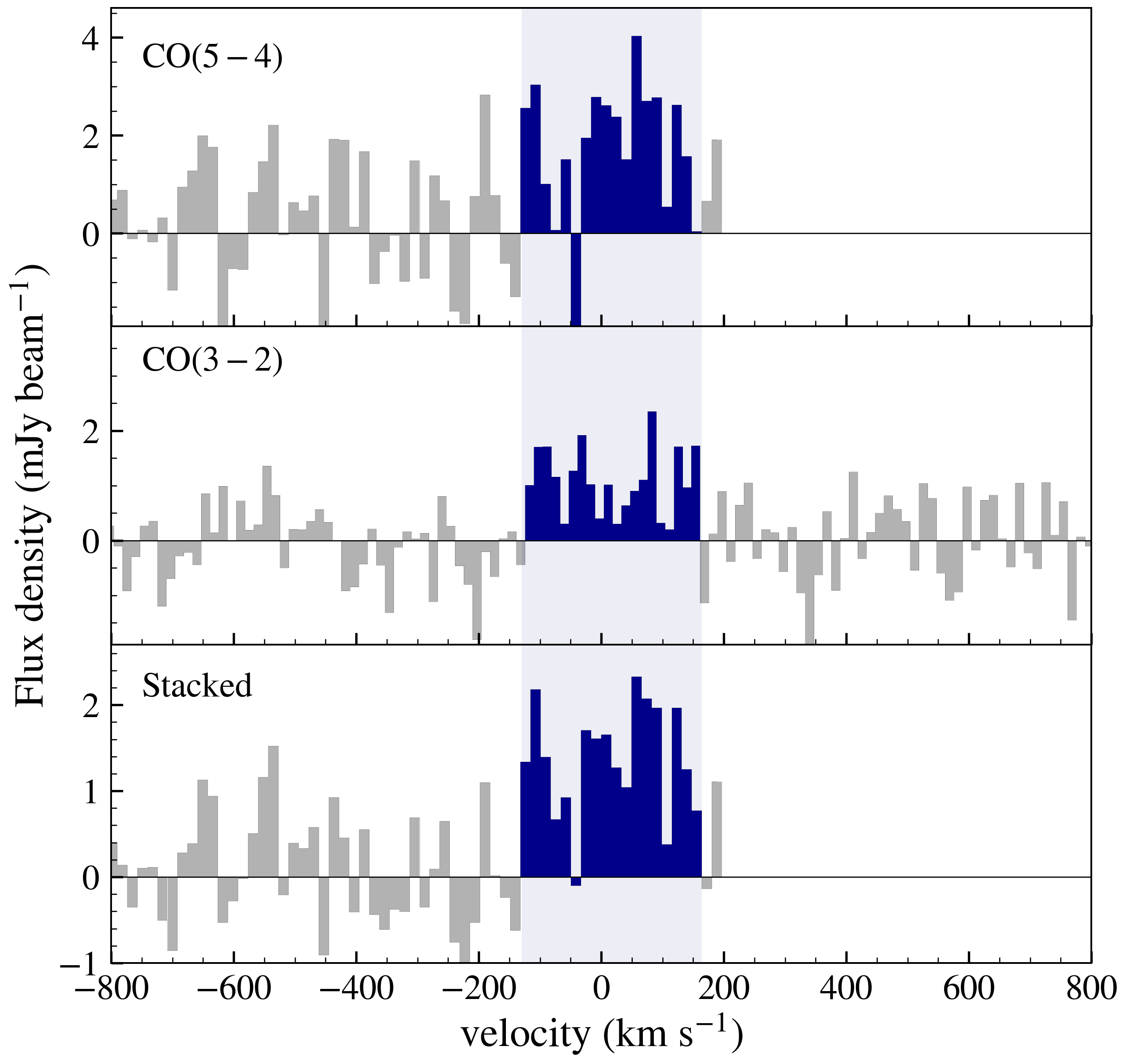}
\centering
\caption{\emph{Left:} HST/WFC3 F160W image for EGS141912 \citep{momcheva2015} with integrated CO($3-2$) and CO($5-4$) moment-0 emission shown as the blue and pink contours respectively. Contours are marked at the $\pm 3,4,5, 6 \sigma$ levels, at 0.1 Jy \kms$ $ beam$^{-1}$ for both maps. \emph{Right:} CO($3-2$) and CO($5-4$) spectra (histograms)from a 1.0'' aperture, as well as their weighted average spectrum. The moment-0 maps were made by summing the channels between $v \in (-130, 164)$ \kms, for which the velocity range was determined based on the detected emission in the rms-weighted stacked spectrum of the CO($3-2$) and CO($5-4$) spectra.}
\label{fig:plt1}
\end{figure*}

%=============================%
% RESULTS
%=============================%

\section{Results}\label{sec:results}

\subsection{CO observations} \label{sec:co54_obs}

We detect CO($5-4$) emission from EGS141912 at $\sim 6 \sigma$ significance, where the moment-0 emission map (\autoref{fig:plt1}) is created by binning the CO($5-4$) line over the same velocities as the CO($3-2$) emission in \citetalias{ag2017} \footnote{We do not fit a 1-D Gaussian to the CO($5-4$) spectral line profile, as the line was observed close to the edge of the spectral band and we lack continuum coverage on one side of the band.}. 

Based on a 2-D Gaussian fitting to the moment-0 map, we find a velocity integrated line flux of $I_{\rm CO(5-4)} = 0.72 \pm 0.12 $ Jy km s$^{-1}$. This corresponds to a line luminosity of $L'_{\rm CO(5-4)} = (1.3 \pm 0.2) \times 10^{10}$ \lu. Both CO($3-2$) and CO($5-4$) spectra are extracted from a circular aperture with radius $1.0''$ centred on the position in Table 2 in order to compare their line profiles, though we caution that this corresponds to a small fraction of the beam for the CO($3-2$) cube, given its $\sim4\times$ coarser spatial resolution (see \autoref{fig:plt1}). We do not detect any continuum emission from EGS141912, giving a $3\sigma$ upper limit of $f_{\lambda} \leqslant 0.1$ mJy at $\lambda_{\rm obs} = 2.2$ mm. 

We also create the rms-weighted average of the CO($3-2$) and CO($5-4$) spectra (\autoref{fig:plt1}) and detect the combined emission at a $\sim 8\sigma$ significance, resulting in an improved $z_{\rm spec} = 3.2185 \pm 0.0002$. The total gas mass has been derived using the CO($3-2$) line (as in \citetalias{ag2017}) and the line luminosities are listed in Table 2. 

\subsection{Radio continuum observations}

We do not detect 9 GHz radio continuum emission from EGS 141912 at the spatial position of the CO emission, and find a $3\sigma$ upper limit of $f_{\rm 9 GHz} \lesssim 3.9$ $\mu$Jy \footnote{We assume that the emission is not spatially resolved in the X-band observations, as it is not resolved in the CO($5-4$) emission, observed with a similar beam size.}. We use this limit to constrain the 1.4 GHz luminosity ($L_{\rm 1.4 GHz}$) as follows 

\begin{equation}
\label{eqn:eqn1}
L_{\rm 1.4 GHz} = \frac{4 \pi D_{L}^{2}}{(1+z)^{1+\alpha}} (\frac{1.4}{\nu_{\rm obs}})^{\alpha}S_{\nu_{\rm obs}}
\end{equation}

where $D_{\rm L}$ is the luminosity distance in metres, $z$ is the source redshift, $\nu_{\rm obs}\sim 9$ GHz, and $\alpha$ is the radio spectral slope of $\alpha = -0.7$ (such that $S_{\nu} \propto \nu^{\alpha}$). This gives a $3\sigma$ upper limit on the 1.4 GHz luminosity of $L_{\rm 1.4 GHz} \lesssim 8.7 \times 10^{23}$ W Hz$^{-1}$. %We compare this to the $L_{\rm IR}$. 

%=============================%
% SED-FITTING
%=============================%

\subsection{SED fitting} \label{ssec:sedfitting}

%=============================%
% TABLE 1 
%=============================%

\begin{deluxetable}{lllll} \tablewidth{0pt}
\tablecaption{Continuum fluxes for EGS141912.}
\tablehead{
\colhead{Telescope}	& \colhead{Band}	& \colhead{$\lambda_{\rm eff}$ ($\mu$m)}	&	\colhead{Flux (mJy)}	& \colhead{Ref}}
\startdata
CFHTLS		& $r'$      & 0.63      & $(3.6 \pm  1.4) \times 10^{-5}$	& (1)	\\ 
			& $i'$      & 0.77      & $(9.0 \pm  1.8) \times 10^{-5}$ 	& (1)   \\ 
\emph{HST} 	& F606W 	&  0.59     & $(5.3 \pm  2.2) \times 10^{-5}$ 	& (1)   \\ 
			& F814W    	& 0.83      & $(7.7 \pm 3.4) \times 10^{-5}$   	& (1)   \\ 
			& F125W    	& 1.25		& $(1.2 \pm 0.3) \times 10^{-4}$  	& (1)   \\ 
			& F140W    	&  1.39     & $(2.1 \pm 0.4) \times 10^{-4}$  	& (1)   \\ 
			& F160W    	&  1.54     & $(2.8 \pm 0.3) \times 10^{-4}$ 	& (1)   \\ 
\spit 		& IRAC	& 3.6	& $(2.0 \pm 0.3) \times 10^{-3}$    & (2)   \\ 
			& IRAC   	& 4.5   & $(2.5 \pm 0.5) \times 10^{-3}$    & (2)   \\ 
			& IRAC    & 5.8   & $(7.0 \pm 1.0) \times 10^{-3}$ 	& (2)   \\ 
			& IRAC    & 8.0   & $(2.6 \pm 1.2) \times 10^{-3}$ 	& (2)   \\ 
			& MIPS     	& 23.7  & $(5.0 \pm 0.7) \times 10^{-2}$  	& (2)   \\ 
\her		& PACS   	& 160   & $(1.8 \pm 0.7) \times 10^{1}$     & (3)   \\ 
NOEMA		& 	     & $2.2 \times 10^{3}$     & $< 0.1$            				& (4)    \\ 
			&	     & $3.7 \times 10^{3}$     & $< 0.3$            				& (5)    \\ 
VLA			& 	     & $3.4 \times 10^{4}$     & $< 3.9 \times 10^{-3}$            	& (5)    \\ 
\enddata
\tablerefs{(1) 3D-HST AEGIS catalog \cite{brammer2012,skelton2014}, (2) \cite{park2010} (3) \cite{oliver2012} (4) \citetalias{ag2017} (5) This work.}
\label{tab:phot}
\end{deluxetable}

To obtain the stellar mass, we adopt the spectral energy distribution (SED) fitting package Code for Investigating GALaxy Emission (CIGALE ; \citealt{burgarella2005, noll2009, serra2011} as described in \citetalias{ag2017} with minor changes (see \autoref{appendix} for more details). We here only use those photometric data points where the emission is detected at SNR $\gtrsim 2$ as well as the upper limits on continuum emission based on our CO observations. The best-fit SED is shown in Figure 2, and the results of the SED fitting as well as all source properties are listed in Table 2. Based on the stellar mass based on the SED fit and gas mass based on the CO($3-2$) line strength, we find a gas mass fraction $f_{\rm gas} = M_{\rm gas}/(M_{\rm gas} + M_{*}) = 0.9 \pm 0.2$. The quoted uncertainty in the gas fraction does not include the systematic uncertainty associated with the stellar mass estimate due to assumptions about the star-formation history ($\sim 30\%$, see \autoref{appendix}), the uncertainties in the CO line luminosity ratio $L'_{\rm CO(3-2)}/L'_{\rm CO(1-0)}$, assumed to be $r_{\rm 31} =  0.42 \pm 0.07$ based on \citet{daddi2015}, or systematic uncertainties in the CO-$\rm H_{2}$ gas mass conversion factor $\alpha_{\rm CO}$ (see \citealt{bolatto2013} for a review).

There are large uncertainties associated with the $L_{\rm IR}$ for EGS141912. This is best demonstrated in Figure 2, where we compare the best-fit SED from CIGALE to high-$z$ SED templates, both for normal and starburst galaxies \citep{magdis2012}\footnote{\url{http://georgiosmagdis.pbworks.com/w/page-revisions/59019974/SED\%20Templates}}. It is clear that in the absence of photometry sampling the peak of the IR emission, the shape of the SED - and therefore the integrated $L_{\rm IR}$ - is poorly constrained. Physically, this arises because a mixed dust/star system may look identical to a dimmer, dust-free system at optical/UV wavelengths, and the two can be distinguished only using far-IR photometry. This lack of far-IR coverage also results in relatively poorly constrained dust mass obtained through SED fitting, $M_{\rm dust} = (6.4 \pm 4.7) \times 10^{8} M_{\odot}$ (also see \citealt{berta2016}). This corresponds to a gas-to-dust mass ratio of $\delta_{\rm GDR} = 400 \pm 300$, which is higher than but consistent with the expected $\delta_{\rm GDR} \sim 100$ for solar metallicities \citep{leroy2011} within the uncertainties.

Anchoring SED templates for $z\sim 3$ main-sequence galaxies to the 24\mum flux \citep{magdis2012}, we infer an IR luminosity of $L_{\rm IR}^{\rm MS} = (2.1 \pm 0.3) \times 10^{12} L_{\odot}$. To get an upper limit on the $L_{\rm IR}$, we fit the upper limit on the NOEMA 2mm continuum flux with a Modified Blackbody function combined with a power-law mid-IR emission \citep[see][for details]{pavesi2016}. We here assume an uniform prior on the dust temperature of $T_{\rm dust} = 35 \pm 10$ K (as suitable for $z\sim 3$ galaxies, \citealt{magnelli2014}) and a dust emissivity of $\beta = 1.7 \pm 0.2$ \citep{planck2014}. We find an upper limit of $L_{\rm IR} \lesssim 4.8 \times 10^{12} L_{\odot}$ with a 99.7\% confidence limit. Overall, we treat the $L_{\rm IR}$ as lying between the $L_{\rm IR}^{\rm lower} = 2.1 \times 10^{12} L_{\odot}$ and $L_{\rm IR}^{\rm upper} = 4.8 \times 10^{12} L_{\odot}$. These limits on the $L_{\rm IR}$ are consistent with those derived using the upper limit on the 1.4 GHz luminosity $L_{\rm 1.4 GHz}$ when assuming a redshift-dependent radio-IR correlation \footnote{The evolution of $q_{\rm IR}$ is an open question, with some studies finding a weak redshift evolution \citep{magnelli2015, calistro2017, delhaize2017}, and with others finding differential evolution for for disc- vs spheroid- dominated galaxies \citep{molnar2018}.}\citep{delhaize2017}. We find $q_{\rm IR} \sim 2.2$ for $z\sim 3.2$ (assuming $\alpha = -0.7$) as compared to $q_{\rm IR} \sim 2.6$ for a non-evolving radio-IR correlation \citep[see Fig 3][]{molnar2018}. These correspond to upper limits on the $L_{\rm IR} \lesssim 1.4 \times 10^{12} L_{\odot}$ and $L_{\rm IR} \lesssim 3.4 \times 10^{12} L_{\odot}$, respectively. 

We use the limits on $L_{\rm IR}$ to get limits for the SFR$_{\rm IR} = 1.09 \times 10^{-10} L_{\rm IR}$ \citep{chabrier2003}, finding SFR$_{\rm IR} = 230 - 520 M_{\odot}$ yr$^{-1}$. EGS141912 then has a specific star-formation rate of sSFR$ = 7.6 - 17.4 $ Gyr$^{-1}$ and gas depletion timescales of $\tau_{\rm dep} = 1.1 - 0.5 $ Gyr. The sSFR is thus $0.9 - 2.1 \times$ sSFR$_{\rm MS}$, where sSFR$_{\rm MS}$ is the sSFR expected from a galaxy lying on the MS at $z\sim 3.2$ \citep{speagle2014, tacconi2018}. EGS141912 is therefore consistent with the MS at $z\sim 3.2$. 

%=============================%
% FIGURE 2 
%=============================%

\begin{figure}
\label{fig:plt22}
\includegraphics[width=0.48\textwidth]{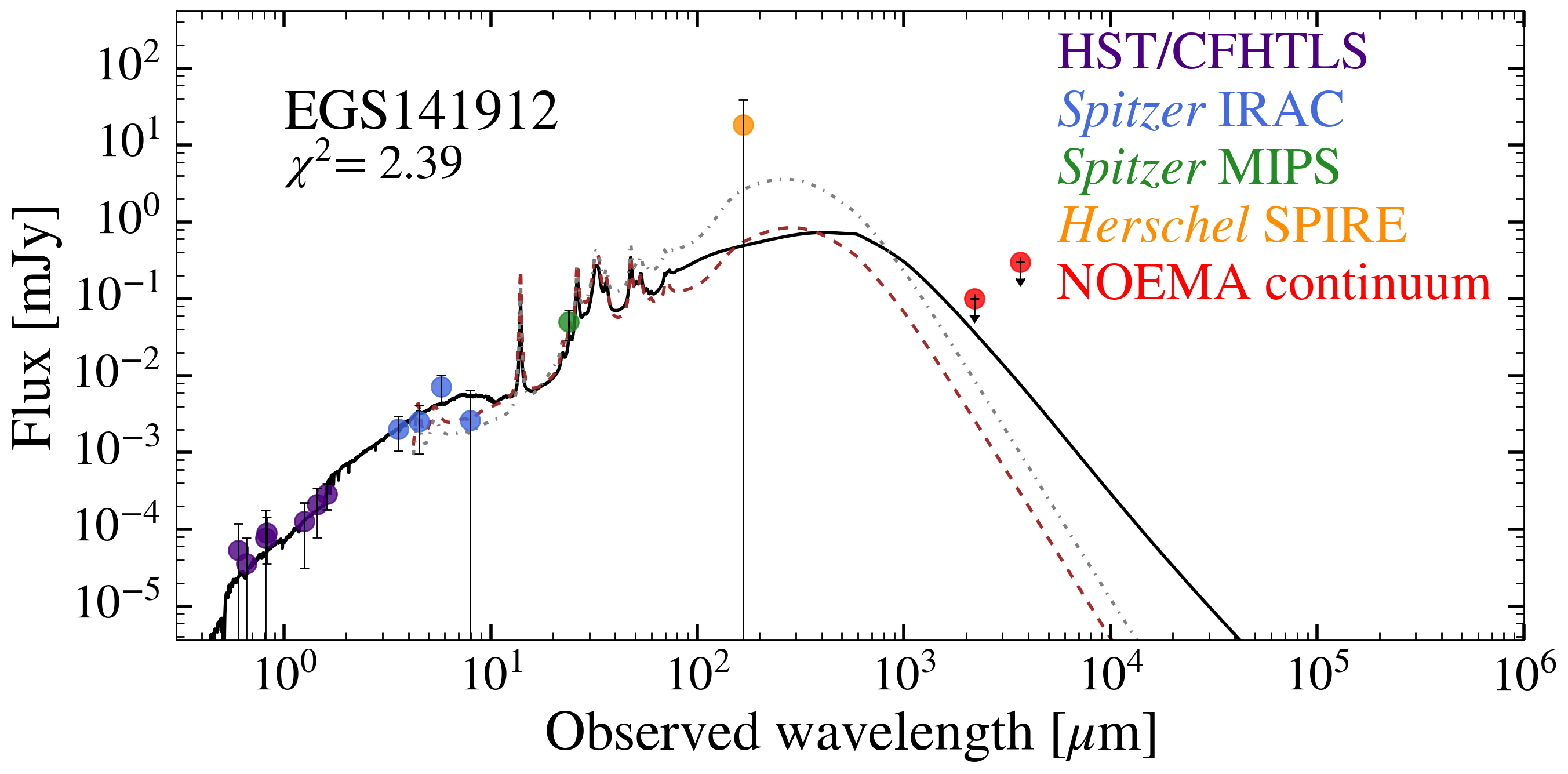}
\caption{Results from SED-fitting for EGS141912 using CIGALE. The colored points represent the observed photometry, listed in \autoref{tab:phot}, and the gray line represents the best-fit SED. The far-IR tail of the SED is poorly constrained due to the lack of available photometry at those wavelengths. The dashed and dash-dotted lines show the fit using \citet{magdis2012} templates for  main-sequence ($z\sim 3$) and starburst galaxies.}
\end{figure}

%=============================%

%=============================%
% TABLE 2 
%=============================%

\begin{table}[]
\label{tab:properties}
\centering 
\caption{Physical properties of EGS141912}
\begin{tabular}{lr}
\hline  
RA, Dec (J2000) 	& $14^h19^m12.0^s +52^d49^m24^s$ \\
$z_{\rm CO}$ 		& $3.2185  \pm 0.0002$ \\
$L_{\rm IR}$ 		& $2.1 - 4.8 \times 10^{12}$ $L_{\odot}$\\
$L'_{\rm CO(3-2)}$ 	&  $ (3.0 \pm 0.5) \times 10^{10}$ K km s$^{-1}$ pc$^{2}$ \\
$L'_{\rm CO(5-4)}$ 	&  $(1.3 \pm 0.2) \times 10^{10}$ K km s$^{-1}$ pc$^{2}$ \\
SFR$_{\rm IR}$ 		& $(230 - 520) M_{\odot}$ yr$^{-1}$  \\
$r_{53}$			& $0.41 \pm 0.10$ \\
$M_{*}$ 			& $(3.0 \pm 0.1) \times 10^{10}$ $M_{\odot}$\\
$M_{\rm gas}$\tablenote{We adopt the total molecular gas mass of $ M_{\rm gas} = (2.6 \pm 0.4) \times 10^{11} M_{\odot}$, as reported in \citetalias{ag2017}, calculated using $L'_{\rm CO(3-2)}$, assuming a line luminosity ratio of $r_{31} = 0.42$ (the average from the \citealt{daddi2015} sample of $z \sim 1.5$ BzK galaxies) and a CO-H$_{2}$ gas mass conversion factor $\alpha_{\rm CO}= 3.6 $ \acou, suitable for main-sequence galaxies at high redshift \citep{daddi2010, carleton2017}.}
 		& $(2.6 \pm 0.4) \times 10^{11}$ $M_{\odot}$\\
$M_{\rm dust}$ 	& $(6.4 \pm 4.7) \times 10^{8} M_{\odot}$\\ 
$\delta_{\rm GDR}$ 	& $400 \pm 300$\\
$f_{\rm gas}$		& $0.9 \pm 0.2$\\ 
sSFR 			& $7.6 - 17.4$ Gyr$^{-1}$ \\ 
\hline 
\end{tabular}
\end{table}

%=============================%
% DISCUSSION
%=============================%

\section{Discussion}\label{sec:discussion}

\subsection{CO excitation at $z\sim 3$}

In general, the CO excitation (measured by line luminosity ratio between high-$J$ and low-$J$ CO lines) is expected to increase at higher-$z$ due to the increased dust temperature \citep{magdis2012}, and potentially due to higher dense gas fractions and star formation efficiencies \citep[e.g.][]{daddi2010, scoville2016}. Such a warm, highly excited molecular gas component is also expected based on simulations of gas excitation and feedback at higher redshifts \citep[e.g.][]{narayanan2014, bournaud2015}. We here quantify the CO excitation in EGS141912 using the CO($3-2$) and CO($5-4$) line detections. For EGS141912, we find a line luminosity ratio of $L'_{\rm CO(5-4)}/L'_{\rm CO(3-2)} = 1.3 \pm 0.2 / 3.0 \pm 0.5 = 0.41 \pm 0.10$. This is slightly lower than but consistent with the excitation observed for BzK galaxies ($L'_{\rm CO(5-4)}/L'_{\rm CO(3-2)} = 0.53 \pm 0.19$; \citealt{daddi2010, daddi2015}), and is lower than the observed excitation in submillimetre galaxies (SMGs ; $r_{53} = 0.61 \pm 0.20$; \citealt{bothwell2013}). 

The star-formation efficiency in EGS141912 ($L_{\rm IR}/M_{\rm gas} \sim (8.1 - 18.4) L_{\odot}/M_{\odot}$) is also consistent with those observed in BzK galaxies ($L_{\rm IR}/M_{\rm gas} \sim (13 \pm 3) L_{\odot}/M_{\odot}$ \citealt{daddi2015}). 

\subsection{The CO-$L_{\rm IR}$ correlation}

%=============================%
% FIGURE 3
%=============================%

\begin{figure*}
\includegraphics[width=0.45\textwidth]{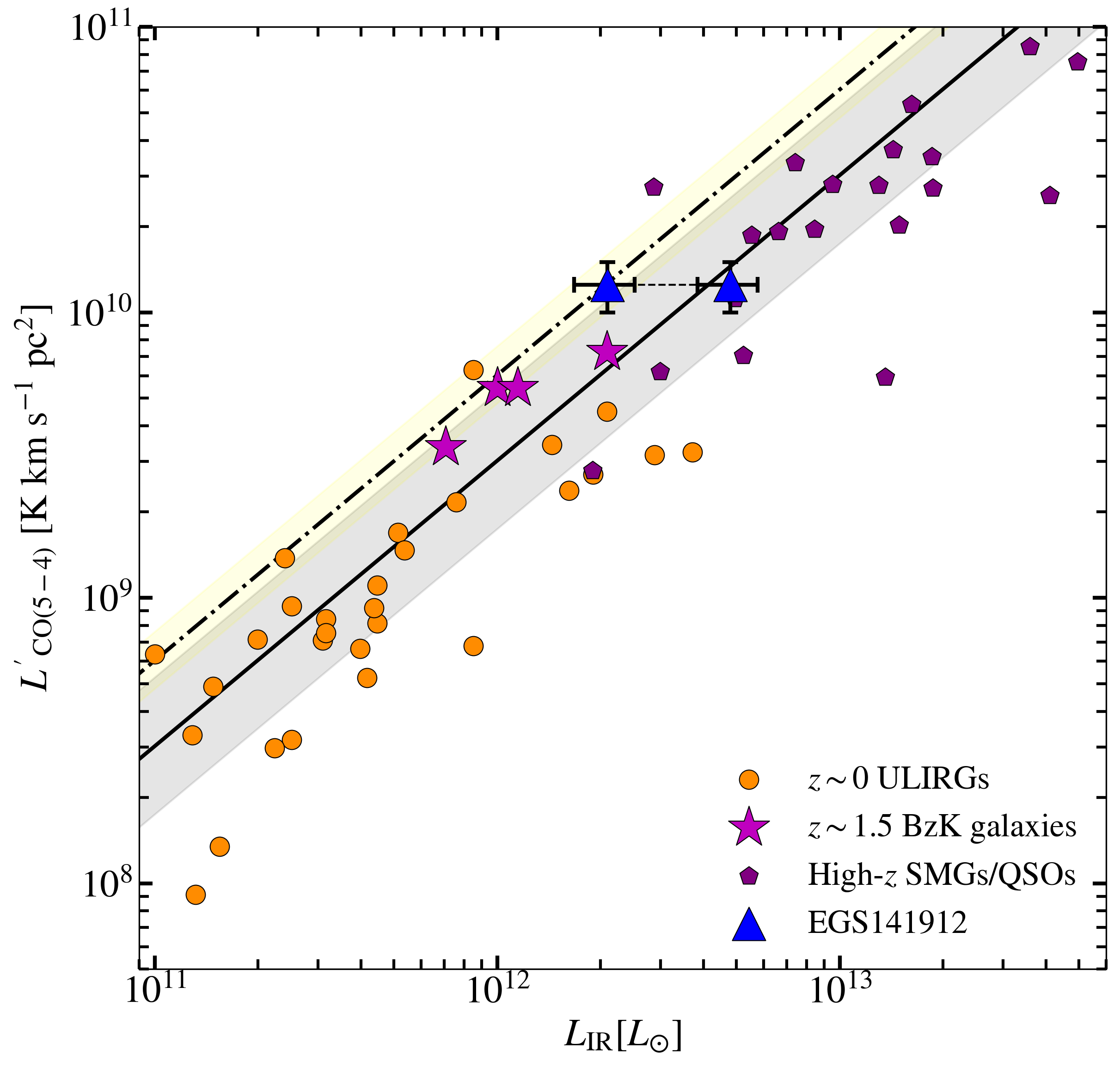}
\includegraphics[width=0.45\textwidth]{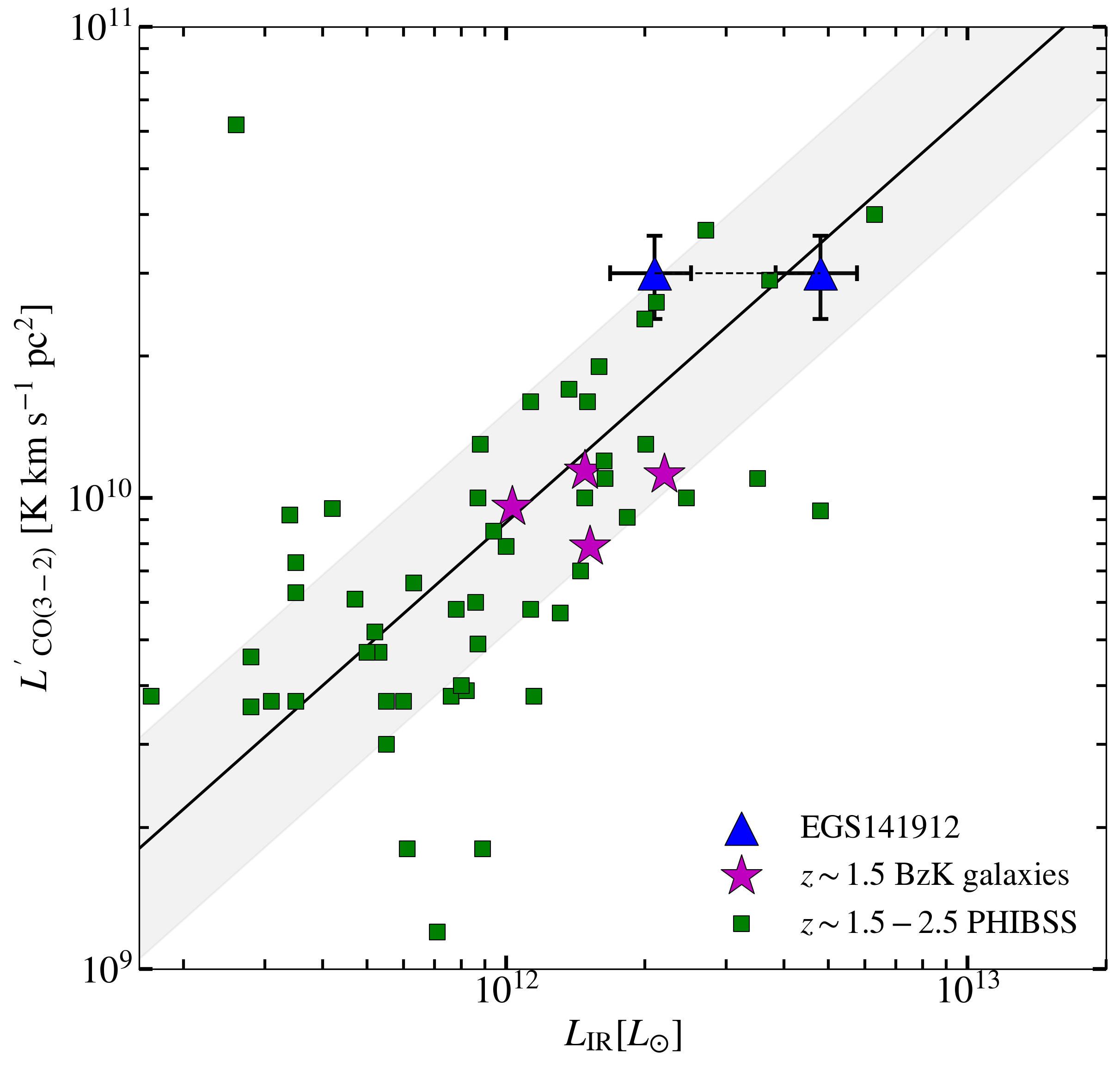}
\centering
\caption{Left: A comparison of $L'_{\rm CO(5-4)}$ vs $L_{\rm IR}$ for EGS141912 for galaxies including local (U)LIRGs, BzK galaxies, and high-$z$ SMGs and QSOs. Right: $L'_{\rm CO(3-2)}$ vs $L_{\rm IR}$ for star-forming galaxies observed at high-$z$ \citep{tacconi2013, daddi2015}. The solid line and the dash-dotted line in the left panel show the best fit relations for all galaxies, and that for local spirals and LIRGs, respectively \citep{daddi2015}. The solid line in the right panel shows the best fit for all galaxies including local spirals, (U)LIRGs, as well as high-$z$ SMGs and QSOs \citep{sharon2016}, assuming SFR $ = 1.09 \times 10^{-10} L_{\rm IR}$ based on a Chabrier IMF \citep{chabrier2003}. The shaded regions around each line show the $1\sigma$ deviation, assuming a constant slope.}
\label{fig:plt3}
\end{figure*}

CO($5-4$) emission is a tracer of warm and dense molecular gas. $L'_{\rm CO(5-4)}$ has been observed to correlate linearly with star formation rates and with $L_{\rm IR}$ in galaxies ranging from local spirals and (U)LIRGs  to high-$z$ star-forming galaxies, SMGs and QSOs \citep[e.g.][]{liu2015, daddi2015, yang2017}. This correlation is somewhat indirectly driven, as the CO emission arises from warm molecular gas, potentially partially heated by mechanical feedback and winds from star-forming regions. A similar correlation also exists between the $L'_{\rm CO(3-2)}$ and the $L_{\rm IR}$ (see \autoref{fig:plt3}). The observed $L'_{\rm CO(5-4)}$ and $L'_{\rm CO(3-2)}$ are consistent with these relations within the scatter.

%===============================%

\subsection{Evolution of the cosmic gas fraction}

%=============================%
% FIGURE 4
%=============================%

\begin{figure}
\includegraphics[width=0.49\textwidth]{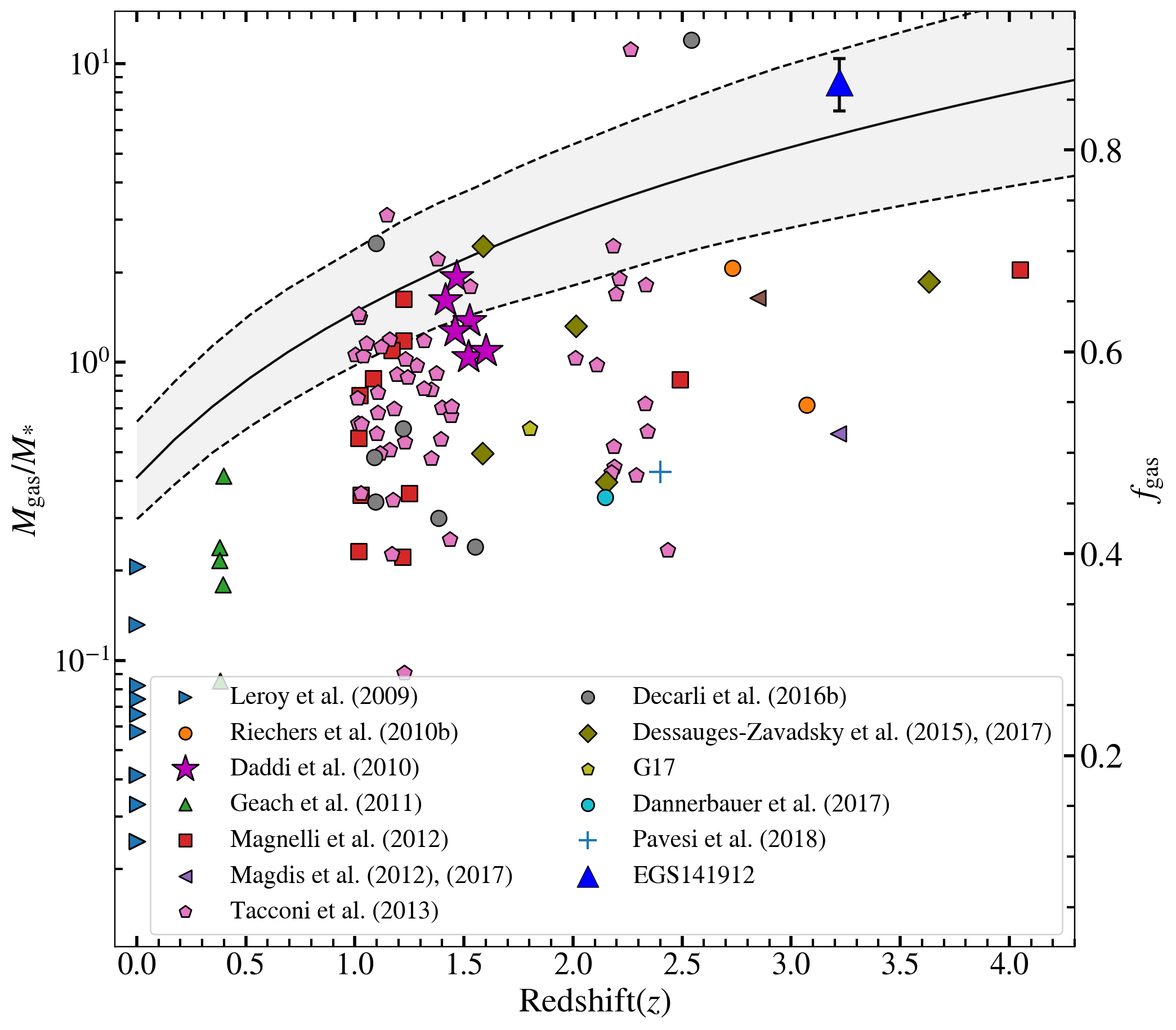}
\caption{The ratio of molecular gas mass to stellar mass (calculated using an \aco $\sim 3.6$ \acou$ $ for all sources) adapted from \citet{carilli2013}. Previous observations are from \cite{leroy2009, riechers2010a, daddi2010, geach2011, magnelli2012, magdis2012, tacconi2013, dz2015, decarli2016a, dz2017, dannerbauer2017, ag2017, pavesi2018}. The black line shows the scaling relation between $f_{\rm gas}$ and $z$, assuming a stellar mass of $M_{*} = 3 \times 10^{10} M_{\odot}$ and $\rm sSFR = 2 \rm sSFR_{\rm MS}$; the shaded regions show the 99.7\% confidence regions. EGS141912 is consistent with an increasing gas fraction at $z \gtrsim 3$.}
\label{fig:plt4}
\end{figure}

The gas fraction $f_{\rm gas}$ in galaxies is a function of $M_{*}$, sSFR/sSFR$_{\rm MS}$, and $z$, with an increasing gas fraction at higher redshift, lower $M_*$, and in galaxies having lying above the MS \citep[e.g.][]{bouche2010, dave2011b, saintonge2011a,saintonge2012,saintonge2017, scoville2017, tacconi2018}. We use the function for this evolution given by \citet{scoville2017}:

\begin{equation}
\begin{aligned}
    f_{\rm gas} = (1.0 + (1.41 \pm 0.18) \times(1.0 + z)^{-1.84 \pm 0.14} \\  
    \times (\rm sSFR/sSFR_{\rm MS})^{-0.32 \pm 0.06} \\    
    \times (M_{*}/10^{10}M_{\odot})^{0.70 \pm 0.04})^{-1}. \\
\end{aligned}
\end{equation}

We compare this against the gas fraction obtained for EGS141912 in \autoref{fig:plt4}. For a MS galaxy at $z\sim 3.2$ with a stellar mass of $M_* = 3 \times 10^{10} M_{\odot}$, the expected gas fraction is $f_{\rm gas} = 0.82$ for $\rm sSFR/sSFR_{\rm ms} = 1.0$, and $f_{\rm gas} = 0.85$ for $\rm sSFR/sSFR_{\rm ms} = 2.0$. EGS141912 shows a gas fraction of $f_{\rm gas} = M_{\rm gas}/(M_{*} + M_{\rm gas}) = 0.9 \pm 0.2$, which falls within a $99.7\%$ confidence interval of the above relation. Similarly high gas fractions have been found in two MS galaxies at $z\sim 2 - 2.5$ \citep{tacconi2013, decarli2016a}, with one showing comparable $M_*$ and $M_{\rm gas}$ to EGS141912, and the other having a significantly lower stellar mass ($M_{*} = 6 \times 10^{9} M_{\odot}$; \citealt{tacconi2013}).

%=============================%
% CONCLUSION
%=============================%

\section{Conclusion}\label{sec:conclusions}

We have presented molecular gas observations of EGS141912, one of the highest redshift unlensed MS galaxies detected in CO to date. Our observations of the CO($3-2$) and CO($5-4$) emission reveal that the gas excitation is consistent with that seen in $z\sim 1.5$ BzK galaxies, although toward the low end. EGS141912 also has a similar star formation efficiency as other high-$z$ MS galaxies between $z\sim 1.5-2.5$. We find EGS141912 to be gas-rich, with a gas fraction of $f_{\rm gas} \sim 0.9 \pm 0.2$, which is consistent with scaling relations for the gas fraction of MS galaxies derived using dust-based measurements of the total ISM mass \citep{scoville2017}. The uncertainties on the star formation efficiency and gas fraction for EGS141912 are driven by those on \aco, $L_{\rm IR}$ and the unknown gas excitation, and we need both high spatial resolution observations of the CO($1-0$) emission as well as observations at the peak of the far-IR SED to improve our knowledge of the cold molecular gas, the molecular gas fraction and its star-formation efficiency. EGS141912 lies well within the attained CO sensitivities by blind surveys such as ASPECS-Pilot \citep{decarli2016a, decarli2016b, walter2016} and COLDz \citep{pavesi2018, riechers2019}. 

While most gas-rich galaxies in the universe at $z > 2$ have optical/IR counterparts \citep{tacconi2013, decarli2016a, pavesi2018}, our findings for EGS141912 show that some of the most gas-rich systems would not be \emph{preferentially} selected for targeted CO follow-up studies at high redshift, either based on optical or far-IR selection criteria (e.g. PHIBBS, \citealt{tacconi2013}). Molecular line scan surveys such as COLDz and ASPECS, which by design are ideal for picking up galaxies like EGS141912, thus provide a complementary probe of the distant universe, and thus, significantly contribute towards our understanding of the total cold gas content throughout cosmic history (e.g., \citealt{decarli2016a, riechers2019}).

\acknowledgements

We thank the referee for excellent and helpful comments which have greatly improved the clarity of the work. A.G acknowledges support from the HST grant HST-GO-14938.003-A. D.R. and R.P acknowledge support from the National Science Foundation under grant number AST-1614213 to Cornell University. RP acknowledges support through the grant SOSPA3-008. This work is based on observations carried out under project number W16DR with the IRAM NOEMA Interferometer. IRAM is supported by INSU/CNRS (France), MPG (Germany) and IGN (Spain). This study makes use of data from AEGIS, a multiwavelength sky survey conducted with the Chandra, GALEX, Hubble, Keck, CFHT, MMT, Subaru, Palomar, Spitzer, VLA, and other telescopes and supported in part by the NSF, NASA, and the STFC. This work is based on observations taken by the 3D-HST Treasury Program (GO 12177 and 12328) with the NSAS/EST HST, which is operated by the Association of Universities for Research in Astronomy, Inc., under NASA contract NAS5-26555. The National Radio Astronomy Observatory is a facility of the National Science Foundation operated under cooperative agreement by Associated Universities, Inc.

\appendix 

\section{Details of SED modelling} \label{appendix}

We have used CIGALE to model the UV to IR SED of EGS141912. Although CIGALE can estimate a large number of galaxy physical properties (including dust attenuation, dust luminosity, $M_{*}$, SFR and $L_{\rm IR}$), given the lack of far-IR photometry for EGS141912, we do not consider the $L_{\rm IR}$ and SFR estimates to be highly reliable (see \autoref{ssec:sedfitting}). 

The modelling and estimation of uncertainties performed by CIGALE have been discussed in greater detail in \citet{noll2009, boq2018}, but we briefly describe them as follows. CIGALE uses independent modules for modelling star-formation histories (SFHs), stellar emission from different population synthesis models \citep{bruzual2003, maraston2005}, dust attenuation \citep{calzetti2000}, dust emission \citep[e.g.][]{draine2007} and radio emission, which together create an integrated SED template. The code implicitly maintains energy balance between the UV attenuation and dust emission. CIGALE takes a range of parameters for each of these modules as input, and builds a model for each combination of parameters. After the grid of normalized models is computed. The models are scaled and compared against the provided photometry, CIGALE finds a likelihood for each of the models, defined as $e^{-\chi^{2}}$. These likelihoods are used to compute the likelihood-weighted mean of the physical parameters and their likelihood-weighted uncertainties, which are returned as the best-fit parameters.

We here focus on the uncertainties on the stellar mass $M_{*}$. For EGS141912, we find a stellar mass of $M_* = (3.0 \pm 0.1) \times 10^{10} M_{\odot}$, assuming a delayed exponential star-formation history, and the \citet{bruzual2003} stellar population synthesis model. To test how robust $M_{*}$ is to our choice of SFH, we have explored the different possible SFHs allowed by CIGALE - a double exponential, a delayed star-formation, as well as a periodic bursts of star-formation. We find a $\sim 30\%$ variation in $M_{*}$ assuming different models, with $M_{*} = (3.0 \pm 0.1) \times 10^{10} M_{\odot}$ for a delayed exponential SFH, to $M_{*} = (3.9 \pm 0.5) \times 10^{10} M_{\odot}$ for periodic bursts of star formation. Assuming a delayed SFH results in the fit with the lowest reduced $\chi^{2} \sim 2.3$, as compared to $\chi^{2} \sim 2.7$ and $\chi^{2} \sim 3.0$ for double exponential SFH and a periodic SFH, respectively. We therefore assume a delayed SFH for the final best-fit SED.

%\bibliography{nat}

\end{document}